\documentclass[11pt,a4paper]{article}
\usepackage[left=15mm, top=10mm, right=15mm, bottom=20mm, nohead=true, nofoot=true]{geometry}

\usepackage[utf8]{inputenc}
\usepackage{authblk}
\usepackage{indentfirst}
\usepackage{graphicx}
\usepackage{amsmath}
\usepackage{amssymb}
\usepackage{multicol}
\usepackage{bm}
\usepackage{longtable}
\usepackage{pdflscape}
\usepackage{epstopdf}
\usepackage{multirow}
\usepackage{adjustbox}
\usepackage{amsfonts}
\usepackage{ifthen}
\usepackage{fancyhdr}
\usepackage{setspace}
\usepackage{xcolor}
\usepackage[round,authoryear]{natbib}
\usepackage[toc,title,page]{appendix}
\usepackage[hidelinks]{hyperref}
\usepackage{url}

\hypersetup{
    colorlinks=true,
    linkcolor=green,
    citecolor=blue,
    filecolor=magenta,
    urlcolor=cyan,
    pdftitle={Sharelatex Example},
    pdfpagemode=FullScreen,
}

\fancypagestyle{plain}{\chead{\texttt{\textcolor{brown}{Communications of BAO, Vol. ?, Issue ?, 20??, pp. ???-???}}}}
\title{\textbf{Mass Loss Rates of Evolved Stars in Isolated Dwarf Galaxies}}

\author[1]{Narges Rostami }
\author[2,3]{Hamidreza Mahani }
\affil[1]{\scriptsize Department of Physics, Faculty of Physics, Alzahra University, Vanak, 1993891176, Tehran, Iran}
\affil[2]{\scriptsize School of Astronomy, Institute for Research in Fundamental Sciences (IPM), P.O. Box 1956836613, Tehran, Iran}
\affil[3]{\scriptsize Iranian National Observatory (INO), Institute for Research in Fundamental Sciences (IPM), P.O. Box 1956836613, Tehran, Iran}

\begin{document}
\pagestyle{empty}
\newpage
\pagestyle{fancy}
\label{firstpage}
\date{}
\maketitle

\begin{abstract}

We derive the mass-loss rates of asymptotic giant branch (AGB) stars in the isolated dwarf irregular galaxy NGC~6822 using \texttt{SEDust}, an automated pipeline that fits spectral energy distributions (SEDs) to estimate stellar mass-loss and dust-production rates. The total mass-loss rate of the AGB population is $(3.7 \pm 1.1) \times 10^{-3}$\,M$_\odot$\,yr$^{-1}$. For a gas-to-dust ratio of 200, this corresponds to a dust injection rate of $2.0 \times 10^{-5}$\,M$_\odot$\,yr$^{-1}$. If sustained over 10\,Gyr, this rate would yield $\approx 2 \times 10^{5}$\,M$_\odot$ of dust, which is lower than the dust mass of $2.7 \times 10^{5}$\,M$_\odot$ estimated from \textit{Herschel} observations. AGB stars alone therefore cannot account for the present-day dust reservoir of NGC~6822, and additional sources or processes, such as dust growth in the interstellar medium, are required.

\end{abstract}

\section{Introduction}

Mass loss from evolved stars plays a key role in stellar evolution and in the chemical enrichment of galaxies, as it returns gas and dust to the interstellar medium \citep{Javadi2013,Abdollahi2023, Mahani2026}. Isolated dwarf galaxies, with relatively simple evolutionary histories and weak gravitational interactions, offer favorable environments for studying this process \citep{Zonoozi19, Mahani21, Gholami25, Mahani23}. The nearby isolated dwarf irregular galaxy NGC~6822 lies at a distance of about $490 \pm 40$\,kpc \citep{Lee1993}, has a stellar mass of $5 \times 10^{7}$\,M$_\odot$, and a metallicity of [Fe/H]~$\approx -1.0$ \citep{Gallart1996, Mahani_Farsi}. It is therefore an ideal target for investigating the mass-loss rates of evolved stars and their contribution to dust production. NGC~6822 has maintained a nearly constant star formation rate over most of the past 11\,Gyr \citep{Clementini2003}. About 3\,Gyr ago, however, it entered a starburst phase \citep{Tolstoy2001}, and its star formation rate has continued to increase over the last 100--200\,Myr \citep{Gallart1996,Khatamsaz2026}. This prolonged star formation implies that the chemical enrichment of its interstellar medium has been continuous, with many stellar generations contributing. Consequently, NGC~6822 hosts a rich population of evolved stars, in particular asymptotic giant branch (AGB) stars.

Asymptotic giant branch stars and long-period variables (LPVs), which represent a late evolutionary stage of low- to intermediate-mass stars ($\sim$0.8--8\,M$_\odot$; \citet{Hofner2018,Javadi2011UK2,Javadi2011,Alizadeh2026,Mina2026,Abdollahi2026,Saremi2021,Saremi2020,Yuan2018,Aghdam2024}), are among the main sources of dust returned to the interstellar medium. During this stage, AGB stars lose a large fraction of their mass through strong stellar winds. The winds are driven by two mechanisms: stellar pulsations, which lift material from the outer layers into the cool extended atmosphere, and radiation pressure on dust grains that condense there, which accelerates the dust and drags the gas outward. Mass-loss rates of AGB stars can reach up to $\sim$ $10^{-4}$\,M$_\odot$\,yr$^{-1}$ (\citet{Hofner2018,Javadi2022,Boyer2025,Abdollahi2023and}). The composition of the dust ejected by AGB stars depends on the surface carbon-to-oxygen (C/O) ratio. In oxygen-rich stars (C/O~$<1$, M-type), the dust consists mainly of silicates and aluminum oxide, whereas in carbon-rich stars (C/O~$>1$, C-type), it is dominated by carbonaceous grains, such as amorphous carbon and silicon carbide (\citet{Boyer2025,Abdollahi2023and,Goldman2019,Gavetti2026}).

In this study, we investigate the mass-loss rates of AGB stars in NGC~6822 and estimate their dust production rates using the method proposed by \citet{Javadi2013}. To this end, we combined optical and near- to mid-infrared photometric catalogs from UKIRT, Spitzer, and JWST to construct a multi-band photometric catalog of AGB stars and LPVs in the galaxy. Their spectral energy distributions (SEDs) were modeled with the radiative transfer code \texttt{DUSTY}, run within the automated \texttt{SEDust} pipeline \citep{Mahani26, Rostami25, Mahani2026}, to derive mass-loss rates and dust-shell properties for the carbon-rich and oxygen-rich populations.

\section{Data}

Our primary catalog of AGB stars in NGC~6822 is taken from \citet{Hirschauer2020}, who identified evolved and dusty stars using near- and mid-infrared photometry. The dataset combines observations from the United Kingdom Infrared Telescope (UKIRT) in the $J$, $H$, and $K$ bands \citep{Sibbons2012} with Spitzer Space Telescope observations in the IRAC 3.6, 4.5, 5.8, and 8.0\,\textmu m bands and the MIPS 24\,\textmu m band \citep{Khan2015}. The catalog contains 1610 AGB stars, of which 1050 are oxygen-rich and 560 are carbon-rich.

To obtain AGB star magnitudes at additional wavelengths, we cross-matched our primary catalog with several other datasets. The first is the optical catalog of \citet{Letarte2002}, who observed NGC~6822 with the CFH12K camera. Using a four-band photometric technique ($R$, $I$, CN, and TiO filters), they identified 904 carbon stars in the region centered on the galaxy. We found 410 carbon stars in common with our primary catalog, which allowed us to add the $R$ and $I$ bands to our photometric data. We also used observations from the James Webb Space Telescope (JWST) to obtain AGB star magnitudes at further infrared wavelengths. The central and active star-forming regions of NGC~6822 were observed with two JWST instruments: NIRCam in the F115W, F200W, F356W, and F444W filters, and MIRI in the F770W, F1000W, F1500W, and F2100W filters.

\begin{figure}[ht]
\centering
\includegraphics[width=0.5\textwidth]{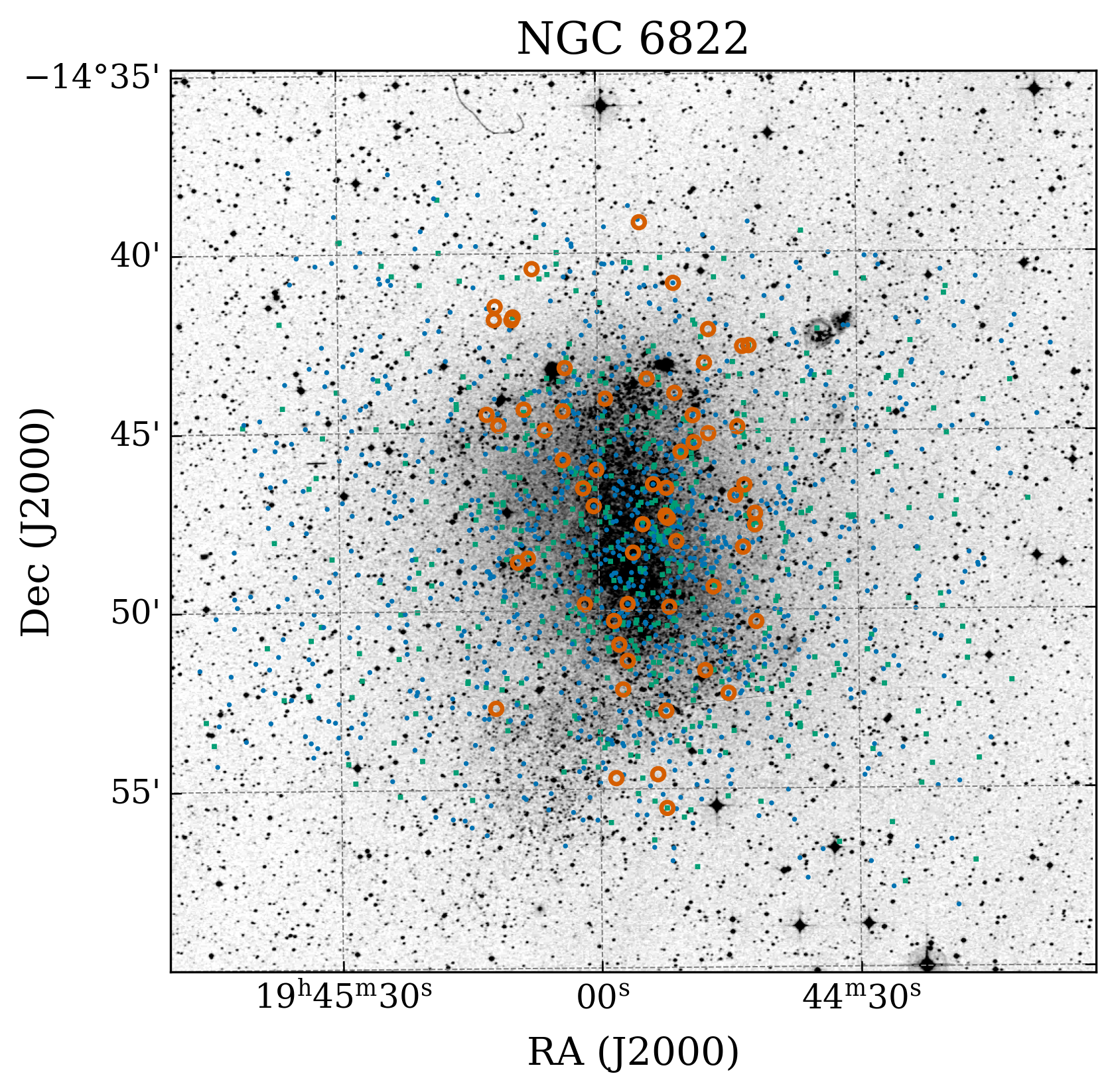}
\caption{Spatial distribution of the evolved stars in NGC~6822. The blue circles show the AGB stars from \citet{Hirschauer2020}, the orange circles show the LPVs from \citet{Whitelock2013}, and the green squares show the carbon stars from \citet{Letarte2002}}
\label{fig:AGB_distrb}
\end{figure}

Finally, we cross-matched our AGB sample with the Mira catalog of \citet{Whitelock2013} and found 59 Miras in our primary catalog, of which 50 are carbon-rich and 9 are oxygen-rich. Mira variables are a subgroup of LPVs associated with the late stages of AGB evolution, with periods typically ranging from about 150 to 1500 days. Cross-matching with these additional catalogs enabled us to compile photometric measurements in up to 18 wavelength bands. Figure~\ref{fig:AGB_distrb} shows the spatial distribution of our AGB and LPV sample in NGC~6822.

\section{Method}

\begin{figure}[ht]
\centering
\includegraphics[width=0.7\textwidth]{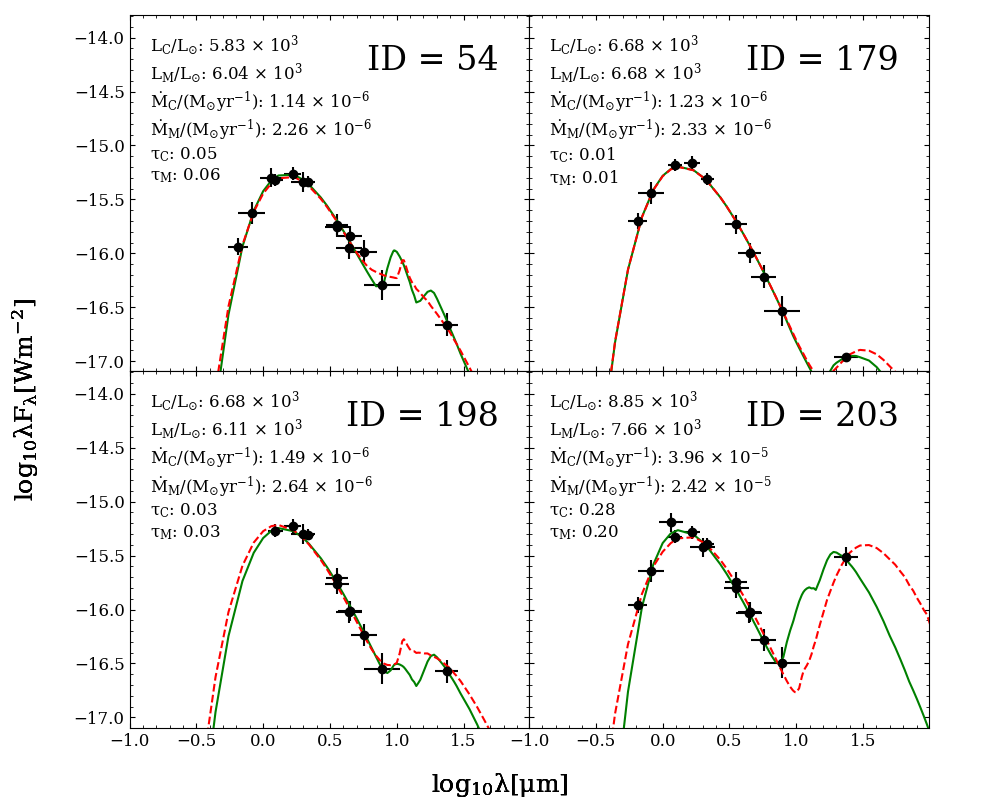}
\caption{Examples of best-fit SEDs for AGB stars. The black points with error bars show the observed photometry. The solid green and dashed red lines show the best-fitting models for M-type and C-type AGB stars, respectively.}
\label{fig:SEDs}
\end{figure}

In this study, we calculated the mass-loss rates of the 1610 AGB stars in our catalog by modeling their SEDs with the  \texttt{DUSTY} code \citep{Ivezic1997}. \texttt{DUSTY} solves the problem of radiation transport through a dusty circumstellar shell. Its key input parameters include the stellar effective temperature, the dust temperature, the optical depth and the chemical composition of the dust. The photometric magnitudes were first converted into fluxes, which were then passed to the \texttt{SEDust} pipeline \citep{Mahani2025,Mahani2026} to find the best-fitting model for each source. \texttt{SEDust} compares \texttt{DUSTY} model outputs with the observed SEDs using $\chi^2$ minimization over a grid of 25\,000 models, which allows reliable fitting of both carbon-rich and oxygen-rich AGB stars. For each star, the pipeline provides key physical parameters, including luminosity, optical depth and mass-loss rate, together with the corresponding best-fit SED plot (Figure~\ref{fig:SEDs}).

\section{Results}

The mass-loss rates of all oxygen-rich and carbon-rich AGB stars were derived by fitting their SEDs with the \textsc{SEDust} pipeline. The total (gas and dust) mass-loss rate of the 1610 AGB stars is $3.7 \times 10^{-3}$\,M$_\odot$\,yr$^{-1}$. Adopting a gas-to-dust ratio of 200 \citep{Ivezic1997}, we obtain a dust injection rate of $\approx 2 \times 10^{-5}$\,M$_\odot$\,yr$^{-1}$. If this rate had remained constant over 10\,Gyr, AGB stars would have returned $\approx 2 \times 10^{5}$\,M$_\odot$ of dust to the ISM. This is lower than the dust mass of $\approx 2.7 \times 10^{5}$\,M$_\odot$ that \citet{Galametz2010} derived from \textit{Herschel} observations of the central region of NGC~6822. AGB stars alone therefore cannot account for the observed dust mass, and other processes, such as dust growth in the ISM, probably contribute.

The ratio of the AGB mass-loss rate to the current star formation rate (SFR $\approx 10^{-2}$\,M$_\odot$\,yr$^{-1}$; \citealt{Efremova2011}) is $\dot{M}_{\mathrm{AGB}}/\mathrm{SFR} \approx 0.37$. Mass returned by AGB stars alone therefore cannot balance the gas consumed by star formation. Part of the gas required to sustain star formation may originate from other sources, such as massive stars or accretion from the surrounding environment.

\begin{figure}[ht!]
  \centering
  \begin{minipage}{0.48\textwidth}
    \centering
    \includegraphics[width=\linewidth]{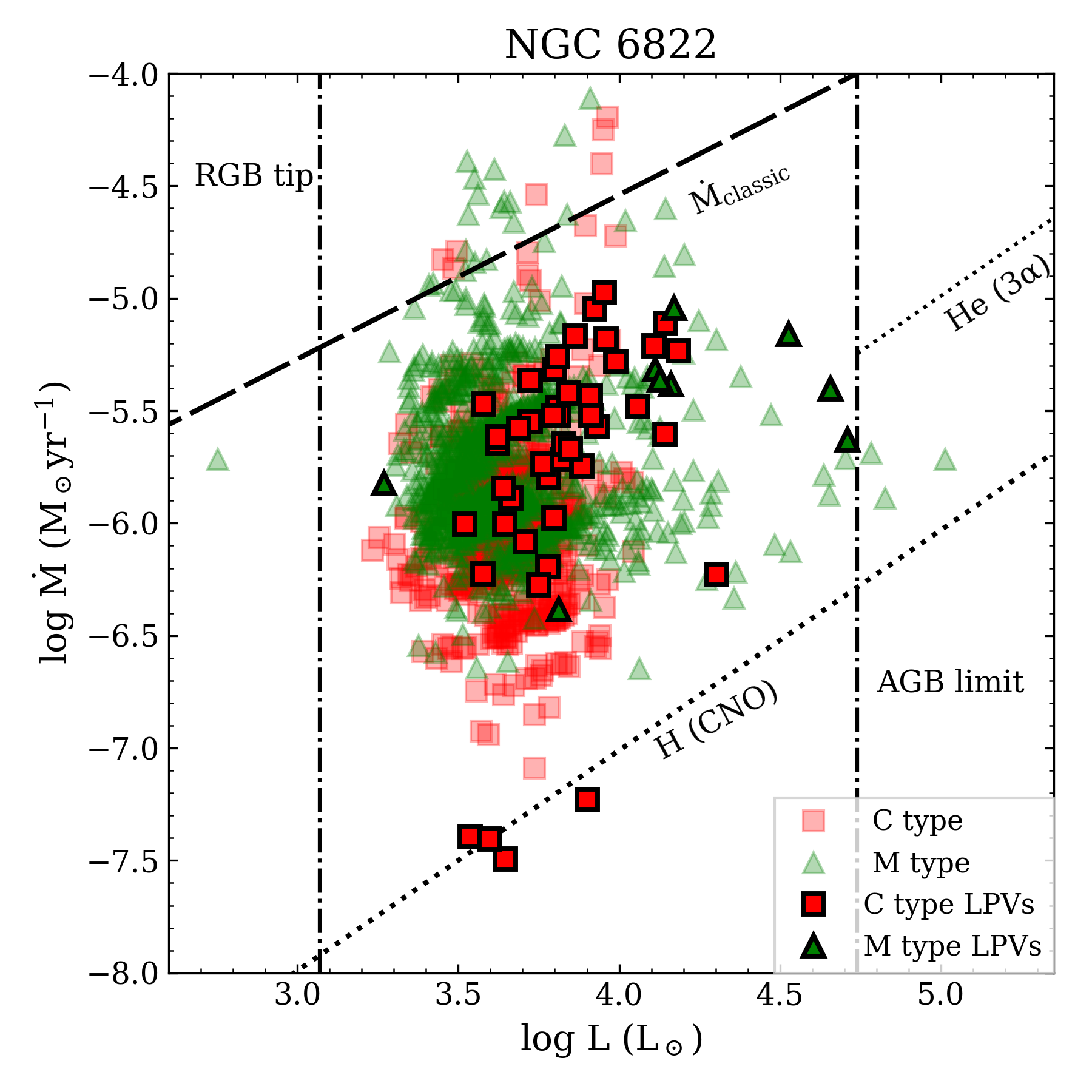}
    \caption{Mass-loss rates of evolved stars as a function of luminosity. Oxygen-rich stars are shown as green triangles and carbon-rich stars as red squares. LPVs are marked with the same symbols outlined in black. The dashed diagonal line shows the single-scattering limit for dust-driven winds \citep{Javadi2013}.}
    \label{fig:mdot6822}
  \end{minipage}
  \hfill
  \begin{minipage}{0.48\textwidth}
    \centering
    \includegraphics[width=\linewidth]{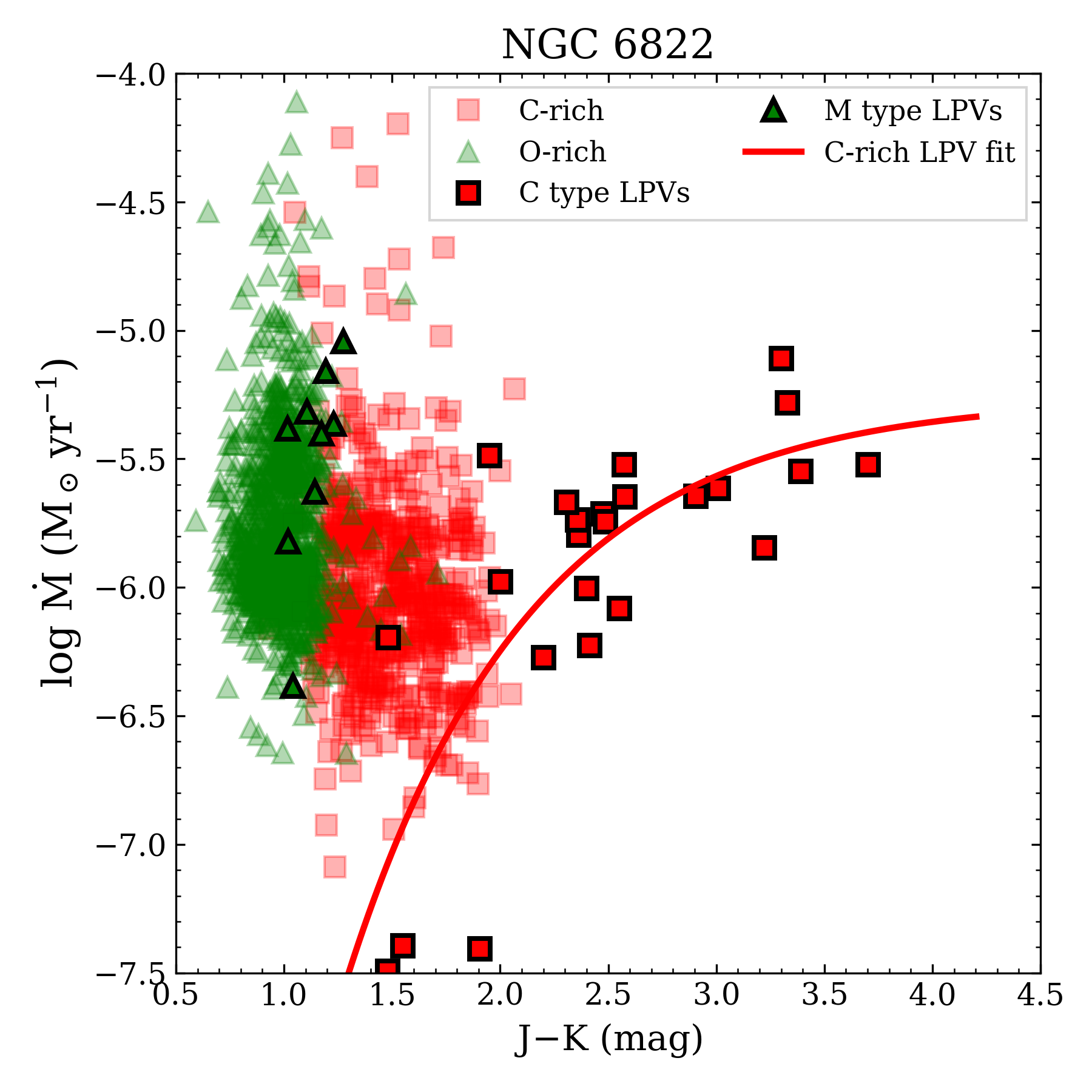}
    \caption{Mass-loss rates of evolved stars as a function of the near-infrared color $J-K$. Oxygen-rich stars are shown as green triangles and carbon-rich stars as red squares. LPVs are marked with the same symbols outlined in black.}
    \label{fig:mdotJK}
  \end{minipage}
\end{figure}

Figure~\ref{fig:mdot6822} shows the mass-loss rate as a function of luminosity for the evolved stars in NGC~6822. The mass-loss rate increases with luminosity. A higher luminosity increases the radiation pressure on circumstellar dust grains, which drives the wind and results in a higher mass-loss rate. The dashed diagonal line shows the single-scattering limit for dust-driven winds, as defined by \citet{Javadi2013}. The mass-loss rates of our sample stars are generally consistent with this limit.

Figure~\ref{fig:mdotJK} shows the mass-loss rate as a function of the $J-K$ color for the sample stars in NGC~6822. This color is commonly used as a combined indicator of the effective temperature of the star and the dust content of its circumstellar shell \citep{Javadi2013}. Figure~\ref{fig:mdotJK} shows that the mass-loss rate of the sample stars in NGC~6822 increases with $J-K$ color. As LPVs evolve along the asymptotic giant branch, their surface temperatures decrease and their $J-K$ colors become redder, owing to molecular absorption and circumstellar dust \citep{Javadi2013}. Red $J-K$ colors indicate cool, extended atmospheres where dust can condense. Once dust forms, radiation pressure drives it outward and enhances mass loss, aided by pulsation-induced shocks that lift material to the dust-forming region \citep{Hofner2018}. Stars with higher mass-loss rates are therefore redder, because of their lower temperatures and larger optical depths.

\section{Summary}

We derived the mass-loss rates of 1610 AGB stars (1050 oxygen-rich and 560 carbon-rich) in the isolated dwarf irregular galaxy NGC~6822. We combined near- and mid-infrared photometry from UKIRT, Spitzer and JWST which gave up to 18 photometric bands per star. Their SEDs were fitted with \texttt{DUSTY} models within the \texttt{SEDust} pipeline.

The main results are as follows:
\begin{itemize}
\item The total (gas and dust) mass-loss rate of the AGB population is $3.7 \times 10^{-3}$\,M$_\odot$\,yr$^{-1}$. For a gas-to-dust ratio of 200, this corresponds to a dust injection rate of $\approx 2 \times 10^{-5}$\,M$_\odot$\,yr$^{-1}$.
\item If this rate had remained constant over 10\,Gyr, AGB stars would have produced $\approx 2 \times 10^{5}$\,M$_\odot$ of dust. This is lower than the dust mass of $\approx 2.7 \times 10^{5}$\,M$_\odot$ derived from \textit{Herschel} observations \citep{Galametz2010}. 
\item The ratio of the AGB mass-loss rate to the current star formation rate is $\dot{M}_{\mathrm{AGB}}/\mathrm{SFR} \approx 0.37$. Mass returned by AGB stars alone therefore cannot balance the gas consumed by star formation, and other sources, such as massive stars or gas accretion, are required.

\end{itemize}


\section*{\small Acknowledgements}
\scriptsize{The authors wish to express their sincere gratitude to the organizers of The International Symposium on Astronomical Surveys and Big Data 3 (ASBD-3) and to the ComBAO collaboration for providing an excellent platform for scientific exchange and collaboration. The valuable discussions and feedback received during the conference greatly contributed to improving the quality of this work. The authors also acknowledge the efforts of the conference committee and local organizers for arranging a well-structured and stimulating meeting that fostered fruitful interactions among participants.}

\scriptsize
\bibliographystyle{ComBAO}
\nocite{*}
\bibliography{V1}

\renewcommand{\thesection}{\Alph{section}.\arabic{section}}
\setcounter{section}{0}
\normalsize

\end{document}